\documentclass{statsoc}

\usepackage[a4paper]{geometry}

\usepackage{amsmath}
\usepackage{enumerate}
\usepackage{natbib}
\usepackage[caption = false]{subfig}
\usepackage{graphicx}
\usepackage{amssymb}
\usepackage{etoolbox}
\usepackage{color}
\usepackage{multirow}
\usepackage{float}

\newcommand{\indep}{\rotatebox[origin=c]{90}{$\models$}}

\title[A Calibrated Sensitivity Analysis for Matched Observational Studies]{A Calibrated Sensitivity Analysis for 
Matched Observational Studies 
with Application to the Effect of Second-hand 
Smoke Exposure on Blood Lead Levels in Children}

\author[B. Zhang and D. S. Small]{Bo Zhang}
\address{Department of Statistics, The Wharton School, University of Pennsylvania, 
Philadelphia, PA, 
U.S.A.}
\email{bozhan@wharton.upenn.edu}
\author[B. Zhang and D. S. Small]{Dylan S. Small}
\address{Department of Statistics, The Wharton School, University of Pennsylvania, 
Philadelphia, PA, 
U.S.A.}

\begin{document}

\begin{abstract}
We conducted a matched observational study to investigate the causal relationship between second-hand smoke and blood lead levels in children. Our first analysis that assumes no unmeasured confounding suggests evidence of a detrimental effect of second-hand smoke. However, unmeasured confounding is a concern in our study as in other observational studies of second-hand smoke's effects. A sensitivity analysis asks how sensitive the conclusion is to a hypothesized unmeasured confounder U. For example, in our study, one potential unmeasured confounder is whether the child attends a public or private school. A commonly used sensitivity analysis for matched observational studies adopts a worst-case perspective, which assumes that in each matched set, the unmeasured confounder is allocated to make the bias worst: in a matched pair, the child with higher blood lead level always attends public school and the other private school. This worst-case allocation of U does not correspond to any realistic distribution of U in the population and is difficult to compare to observed covariates. We proposed a new sensitivity analysis method that addresses these concerns. We apply the new method to our study and find that in order to explain away the association between second-hand smoke exposure and blood lead level as non-causal, the unmeasured confounder would have to be a bigger confounder than any measured confounder.
\end{abstract}

\keywords{Calibration; Matched observational studies; Second-hand smoke; Sensitivity analysis}

\maketitle

\section{Introduction}
\label{section: intro}
\pagenumbering{arabic}
\subsection{An observational study on the effect of second-hand smoke exposure on blood lead levels in children}

Concerns about the health effects of second-hand smoke exposure have been around since at least $1928$ when Sch\"{o}nherr proposed that lung cancers among non-smoking women could be caused by inhalation of their husbands' smoke (\citealp{Smith2003}). A randomized controlled trial is clearly unethical. Over the years, many observational studies have been conducted (e.g. \citealp{Enstrom2003}; \citealp{Mannino2003}; \citealp{Oberg2011}). In this paper, we are concerned with the impact of second-hand smoke on blood lead levels among children. It is widely acknowledged that active tobacco smoking causes higher blood lead levels (\citealp{Shaper1982}; \citealp{Grasmick1985}; \citealp{Mannino2005}). An important public health question arises as whether exposure to second-hand smoke also causes higher blood lead levels, particularly in children. An effect of second-hand smoke on blood lead levels in children would be concerning because children's nervous systems are still developing and high blood lead level is thought to cause consequences such as decreased intelligence, impaired growth, and anemia (\citealp{NRC1993}).

\citet{Mannino2003} studied the relationship between second-hand smoke exposure and blood lead levels in a sample of $5,592$ U.S. children, aged 4-16 years old, and concluded there is strong evidence that second-hand smoke exposure is \emph{associated} with increased blood lead levels. If second-hand smoke exposure were randomized, then association would imply causation, but in fact second-hand smoking was not randomized and any causal conclusion needs to be made with great caution. 

We followed \citet{Mannino2003} in constructing a dataset that included children aged $4$-$16$ years old for whom both serum cotinine levels and blood lead levels were measured in the Third National Health and Nutrition Examination Survey (NHANES III), along with the following variables: race/ethnicity, age, sex, poverty income ratio, education level of the reference adult, family size, number of rooms in the house, and year the house was constructed. The biomarker cotinine is a metabolite of nicotine and an indicator of second-hand smoke exposure (\citealp{Mannino2003}). We excluded active smokers (cotinine level greater than or equal to $15.0$ ng/ml) as in \citet{Mannino2003}, and classified a child as having a high cotinine level (treatment = 1) if the cotinine level is in the third tertile, greater than or equal to $0.563$ ng/ml, and a not high cotinine level (treatment = 0) otherwise.

Two main problems have been raised in making causal inference from observational studies of second-hand smoke's effects: measurement error and confounding (\citealp{Kawachi1996}). Measurement error refers to the inaccuracy in classifying the subjects into the exposed and control groups. This difficulty is largely overcome in our study, because the classification is based on accurate cotinine measurement. A second and perhaps more challenging problem is confounding. Individuals exposed to environmental tobacco smoke usually display adverse profiles in relation to socioeconomic position and health related behaviours (\citealp{Smith2003}). Table \ref{tbl: tableone} summarizes the baseline covariates in the treated and control groups in our study. We see that children exposed to second-hand smoke are in general poorer, younger, live in older houses, and have less rooms in their houses. A direct comparison of treated and control groups is thus not warranted.

\begin{table}
\caption{\label{tbl: tableone}\small Adjusted means and standardized differences of baseline covariates in the treated/control groups before and after full matching: \textbf{EDUCATION} (years of education of the reference adult), \textbf{MALE/FEMALE} (gender), \textbf{NON-WHITE/WHITE} (race/ethnicity), \textbf{POVERTY} (income divided by poverty threshold), \textbf{BEFORE/AFTER 1974} (when the house was built), \textbf{ROOM} (number of rooms in the house), \textbf{FAMILY} (size of family).}
\centering
\resizebox{\columnwidth}{!}{
\fbox{%
\begin{tabular}{c c c c c c c}
  \hline
  &\multicolumn{3}{c}{Before matching} &\multicolumn{3}{c}{After matching}\\
  \hline
 & Control & Treated & Std. diff. & Control & Treated & Std. diff.\\ 
  \hline
EDUCATION & 11.13 & 10.93 & -0.07 &11.16 & 10.93 &-0.08\\ 
  MALE & 0.49 & 0.49 & -0.01  &0.52 &0.51 &-0.01\\ 
  NON-WHITE & 0.73 & 0.71 & -0.05 &0.69 &0.71 &0.03    \\ 
  POVERTY & 1.86 & 1.42 & -0.39  &1.47 &1.42 &-0.04\\ 
  BEFORE 1974 & 0.37 & 0.29 & -0.18  &0.29 &0.29 &-0.01\\ 
  ROOM & 6.00 & 5.67 & -0.20  &5.70 &5.67 &-0.02\\ 
  FAMILY SIZE & 5.08 & 4.92 & -0.08  &4.81 &4.92 &0.06\\ 
  AGE & 9.56 & 9.06 & -0.13 &9.04 &9.06 &0.01\\ 
   \hline
\end{tabular}}}
\end{table}

To control for the measured confounders, we constructed a matched observational study. In a matched observational study, subjects with comparable measured confounders are put into matched sets and inferences are drawn based on comparisons within these matched sets (\citealp{Rosenbaum2002a, Rosenbaum2010}; \citealp{Hansen2004}; \citealp{Stuart2010}; \citealp{Lu2011}; \citealp{Zubizarreta2012}; \citealp{Pimentel2015}). A matched observational study can be analyzed in a model-based or nonparametric way. \citet{Rubin1979} found that the combination of matching and model-based adjustment within matched sets is robust to model misspecification and relatively efficient. 

We performed a full matching (\citealp{Hansen2004}; \citealp{Hansen2006}) on the aforementioned $8$ variables. Table \ref{tbl: tableone} suggests that covariates are well-balanced after matching; in particular, all standardized differences are less than $0.1$ after matching (\citealp{Silber_rosenbaum2016}). The mean blood lead level is $3.27$ $\mu$g/dL. Assuming that there is no unmeasured confounding, the $95\%$ two-sided confidence interval of the effect of second-hand smoke exposure on blood lead level is $(0.71, 0.98)$ $\mu$g/dL using the \textsf{senfmCI} function in the \textsf{R} package \textsf{sensitivityfull} (\citealp{R_pkg_sensitivityfull}) with default settings. This confidence interval is obtained by inverting the test based on Huber's M-statistic for an additive constant treatment effect.

The above analysis suggests evidence for a causal relationship between second-hand smoke and higher blood lead levels in children; however, the analysis relies on assuming ``no unmeasured confounding'', a rather questionable assumption. Although we have put subjects with similar observed covariates in the same matched sets, we are concerned that some unmeasured confounder that we have not matched upon, say some aspect of the socioeconomic status or a genetic variant, might be associated with the cotinine level and the blood lead level, and induce a spurious causal relationship. For instance, one potential unmeasured confounder we are concerned about is whether the child attends a public or private school. Public versus private school could be associated with cotinine exposure because (i) adjusting for reference adult's education and poverty income ratio does not fully adjust for the child's socioeconomic status; consequently, private versus public school may capture residual socioeconomic status of the child after adjusting for reference adult's education and poverty income ratio, and be associated with cotinine level; and (ii) it is conceivable that students attending public schools are at higher risk of cotinine exposure because there are on average more smokers at public schools. Private versus public school could be associated with the potential outcomes for blood lead level because public schools tend to have older buildings and one leading hazard source of lead exposure for US children is lead-based paint which was commonly used in older buildings (\citealp{AAP2005lead}).

\subsection{Sensitivity analysis for matched observational studies: the bounds approach and application to the current study}
\label{intro: literature.review and application}
Sensitivity analysis is one approach to tackling concerns about unmeasured confounding. A sensitivity analysis asks how much of a departure from the no unmeasured confounding assumption would be needed to affect the inferences drawn from an analysis that assumes no unmeasured confounding. A solid causal conclusion should be relatively insensitive to plausible violations of this ``no unmeasured confounding'' assumption. Over the years, many sensitivity analysis methods have been proposed for different causal inference frameworks (\citealp{Cornfield1959}; \citealp{Rosenbaum1983}; \citealp{Gastwirth1998}; \citealp{Imbens2003}; \citealp{McCandless2007};
\citealp{Ichino2008}; \citealp{Hosman2010}; \citealp{carnegie2016assessing}; \citealp{Ding2016}; \citealp{Dorie2016}; \citealp{deLuna2018};
\citealp{fogarty2019studentized};
\citealp{franks2019flexible}; \citealp{cinelli2020making}). 

In this section, we focus on reviewing a commonly used sensitivity analysis framework for matched observational studies: the bounds approach. We illustrate the bounds approach using the simultaneous sensitivity analysis model proposed by \citet{Gastwirth1998}. \citet{Gastwirth1998} hypothesize the existence of an unmeasured variable $U$ such that there would be no more unmeasured confounding were $U$ observed and matched on. Specifically, they consider the following model:
\begin{equation}
    \begin{split}
    &Y_i^{(0)} \indep Z_i \mid \boldsymbol X_i, U_i \\
&P(Z_i = 1 \mid \boldsymbol X_i, U_i) = \text{expit}\{ \kappa(\boldsymbol X_i) + \lambda U_i\} \\
&P(Y_i^{(0)} = y) = \exp\{\xi(U_i, \boldsymbol X_i) + \phi(y, \boldsymbol X_i) + \delta y U_i\} \\
&0 \leq U_i \leq 1,
\label{model: Gastwirth}    
    \end{split}
\end{equation}
where $\boldsymbol{X}_i$ is a vector of observed covariates, $U_i$ the hypothesized unmeasured confounder, $Z_i$ the treatment status, and $Y_i^{(0)}$ the potential outcome under no treatment. This model contains as a special case that the potential outcome under control follows a normal distribution: 
\begin{gather*}
    P(Y_i^{(0)} = y) = \text{Normal}(\xi(\boldsymbol X_i) + \delta U_i, ~\sigma^2).
\end{gather*}

The simultaneous sensitivity analysis model contains two sensitivity parameters: $\lambda$, which controls the strength of the association between $U$ and the treatment status, and $\delta$, which controls the strength of the association between $U$ and the response. Consider a test statistic $t(\cdot)$ for which we want to compute the tail probability $P(t(\cdot) \geq a)$ under the null hypothesis of no treatment effect. A sensitivity analysis computes the tail probability for a given test statistic $t(\cdot)$, a set of sensitivity parameters $(\lambda,\delta)$, and given values of the unobserved covariates $U$. As the values of $U$ vary, the permutation test yields different $p$-values. \citet{Gastwirth1998} show that for matched pairs, a sharp upper bound on the $p$-value (i.e., the most conservative inference) is obtained by letting the unit with higher response have $U = 1$ and the unit with lower response have $U = 0$. \citet{Rosenbaum1987} considers model
(\ref{model: Gastwirth}) except asserts only the model for treatment $Z_i$ and not the model for the outcome $Y_i^{(0)}$ and shows also that a sharp upper bound on the p-value is obtained by letting the unit with higher response have $U = 1$ and the unit with lower response have $U = 0$. \citet{Gastwirth1998} show that this bound is identical to that obtained from taking $\delta \rightarrow \infty$ in their model.   These bounds are sometimes referred to as Rosenbaum bounds in the literature (\citealp{DiPrete2004}). When the matching design is a full match (\citealp{Hansen2004}) rather than a pair match, \citet{Gastwirth2000} and \citet{Small2009} show how to obtain asymptotically valid bounds using a computationally fast algorithm. 

We followed \citet{Gastwirth1998}'s approach and tested the null hypothesis of no treatment effect for various sensitivity parameters $(\lambda, \delta)$ in our matched design. Sensitivity parameters and the corresponding worst-case $p$-values are summarized in Table \ref{tbl: Gastwirth et al res.}. The null hypothesis of no treatment effect can still be rejected when $(\lambda, \delta) = (0.8, 0.8)$, but not when $(\lambda, \delta)$ grows to $(1.0, 1.0)$.

\begin{table}
\caption{\label{tbl: Gastwirth et al res.}
\citet{Gastwirth1998}'s simultaneous sensitivity analysis applied to the second-hand smoke and blood lead levels study}
\centering
\fbox{%
\begin{tabular}{llllll}
\hline
\hline
&($\lambda$, $\delta$) &&& p-value &   \\ \hline
&(0,0)           &&& 0   &\\
&(0.5, 0.5)      &&& 7.55e-15 &   \\
&(0.8, 0.8)      &&& 6.93e-4 &\\
&(1.0, 1.0)      &&& 0.488  &\\
&(1.2, 1.2)      &&& 0.999   &\\
\hline
\end{tabular}}
\end{table}

\subsection{Limitations of the bounds approach, calibration to observed covariates, and motivation for a new framework}
\label{subsec: limitation of GKR and calibration}
The bounds approach derives bounds on the $p$-value, given the sensitivity parameters, by finding the worst-case allocation of the unmeasured confounder in a finite sample. This approach has two limitations. First, it is conservative for an unmeasured confounder in a superpopulation. Consider one potential unmeasured confounder in our study of whether a child attends a public or private school. Let $U = 1$ if the child goes to the public school and $U = 0$ otherwise. The $p$-value reported in the bounds approach of \citet{Gastwirth1998} (see Table \ref{tbl: Gastwirth et al res.}) corresponds to the following worst-case scenario: in each matched pair, the child with higher blood lead level always goes to the public school ($U = 1$), while the child with lower blood lead level always goes to the private school ($U = 0$). However, this hypothesized binary unmeasured confounder, had it existed, must possess a Bernoulli distribution in the population and as we construct matched pairs from the population, there is always some positive probability that both units have $U$ simultaneously equal to $0$ or $1$, i.e., both kids attend the private school or both the public school. The worst-case distribution of $U$ under which the bounds are computed is conservative and does not correspond to any realistic probability distribution for this hypothesized unmeasured confounder in the population.

The bounds approach also does not enable researchers to perform accurate \emph{calibration}. Calibration refers to methodologies that aid researchers in judging plausibility of the unmeasured confounding in reference to observed covariates, i.e., to calibrate how much unmeasured confounding would make a study sensitive to bias in terms of how it compares to the confounding from observed covariates. Several papers have developed calibration methods for various sensitivity analysis frameworks (\citealp{Ichino2008}; \citealp{Hosman2010}; \citealp{Blackwell2013}; \citealp{Griffin2013}; \citealp{Hsu2013}; \citealp{Dorie2016}; \citealp{Middleton2016}; \citealp{cinelli2020making}). In particular, \citet{Hsu2013} develop a calibration approach for matched observational studies using the bounds approach of \citet{Gastwirth1998}. \citet{Hsu2013} calibrate the unmeasured confounder to the observed covariates by comparing the sensitivity parameters $(\lambda, \delta)$ in the simultaneous sensitivity analysis to the coefficients of the observed covariates obtained from a regression analysis. A problem with this approach is that, as explained in the last paragraph, the binary unmeasured confounder has a deterministic structure under the worst-case allocation that the bounds correspond to and therefore cannot be compared to any observed covariate with a probability distribution in a meaningful way. This limitation motivates us to develop a new sensitivity analysis framework for matched observational studies, and leverage it to assess the robustness of our causal conclusion concerning the effect of second-hand smoke exposure on blood lead level in children.

\subsection{Our contribution}
\label{subsec: contribution}
Our sensitivity analysis framework for matched observational studies departs substantively from the bounds approach in two aspects. First, our framework views matched sets as drawn from a population. The hypothesized unmeasured confounder is modeled as a random variable with a probability distribution, instead of having a deterministic structure as in the worst-case allocation under which the bounds are computed (\citealp{Gastwirth1998}; \citealp{Rosenbaum2010}). We borrow from the literature on model-based sensitivity analysis methods (\citealp{Rosenbaum1983}; \citealt{Imbens2003}; \citealp{carnegie2016assessing}; \citealp{cinelli2020making}; see Supplementary Material A.1 for a brief review), and adapt it to matched observational studies. Extensive simulations demonstrate that by combining matching and model-based adjustment, our method is robust to assorted model misspecifications and can be more powerful compared to the bounds approach. 

Additionally, by endowing the hypothesized unmeasured confounder with a probability distribution in the population, our method allows for calibrating the hypothesized unmeasured confounder to the measured confounders in a more meaningful manner compared to the calibration strategy in \citet{Hsu2013}. \citet{Ichino2008} also developed a sensitivity analysis method for matched observational studies with calibration. Our model and calibration approach differ from those in \citet{Ichino2008} in that \citet{Ichino2008} assume that $U$ is independent of the observed covariates conditional on the treatment and the outcome, whereas we assume $U$ and the observed covariates are unconditionally independent (and consequently may be conditionally dependent). Conditional independence between $U$ and $\mathbf{X}$ adopted as in \citet{Ichino2008} may not hold in some cases, as is discussed in detail in Supplementary Material A.2 from a causal directed acyclic graph (DAG) point of view.

More broadly, our calibration method is distinct from many existing calibration methods in that we adjust for the omission of the unmeasured confounder when performing calibration, while most existing approaches, with the exception of \citet{Hsu2013} and \citet{cinelli2020making}, use the observed statistics of the measured covariates for calibration without adjustment for the omission of the unmeasured confounder. As discussed extensively in \citet{cinelli2020making}, this practice has undesirable properties, mainly because the estimates of how the observed covariates are related to the outcome and the treatment assignment are often themselves affected by the omission of the unmeasured confounder $U$, even when $U$ is assumed to be independent of the observed covariates. 

There are two outputs from our sensitivity analysis method. The first output is a calibration plot that contrasts sensitivity parameters to estimated regression coefficients in magnitude. This calibration plot is further supplemented with a table summarizing the variable importance of the hypothesized unmeasured confounder relative to the measured confounders in confounding the outcome and the treatment assignment. We leverage our proposed sensitivity analysis method to assess the robustness of our primary analysis conclusion, i.e., that second-hand smoke exposure has a detrimental effect on children by elevating their blood lead levels. We find that to explain away the causal conclusion, an unmeasured confounder needs to be associated with blood lead level to a similar extent as the poverty income ratio and at the same time be more associated with cotinine level than any of the eight observed covariates.  

The rest of the paper is organized as follows. In Section $2$, we explain notation and present model specifications. In Section $3$, we re-frame the problem as a missing data problem and solve it using an EM algorithm. Section $4$ discusses the proposed calibration approach and Section $5$ presents simulation results. We apply our methodology to do a sensitivity analysis for the effect of second-hand smoke exposure on a child's blood lead level in Section $6$. Our R package \textsf{sensitivityCalibration} implements the methodology and is available through the Comprehensive R Archive Network (CRAN).

\section{Notation and Framework}
\label{section: notation and model}
We clarify the notation and describe our model in this section. Suppose there are $I$ matched sets, $i = 1,2, ..., I$, each matched for a set of observed covariates $\boldsymbol x$. Within each matched set, there are $n_i$ subjects and let $N = \sum_{i = 1}^I n_i$ be the total number of subjects in the study. In the setting of full matching (\citealp{Rosenbaum2002a}; \citealp{Hansen2004}), each matched set consists of either one treated subject and $n_i - 1$ controls, or one control and $n_i - 1$ treated subjects. If $n_i = 2$ for all $i$, then we have the pair matching.  

Let $ij$ denote subject $j$, $j = 1,2, ..., n_i$, either treated or control, in matched set $i$. Let $Z_{ij} = 1$ if subject $ij$ is treated and $0$ otherwise. Let $\boldsymbol x_{ij}$ denote the observed covariates of subject $ij$. According to the matched design, $\boldsymbol x_{ij} = \boldsymbol x_{ij'}$ for $j,j'$ in the same matched set $i$. To conduct sensitivity analysis, we hypothesize the existence of a hidden bias that comes from an unmeasured confounder $u_{ij}$ associated with each subject $ij$. Following the potential outcome framework (\citealp{Neyman1923}; \citealp{Rubin1974}), we let $Y_{ij}^{(1)}$ and $Y_{ij}^{(0)}$ denote the potential outcome of subject $ij$, under treatment $(Z_{ij} = 1)$ or control $(Z_{ij} = 0)$, respectively. Hence, the observed outcome of subject $ij$ is $Y_{ij} = Z_{ij} Y_{ij}^{(1)} + (1 - Z_{ij}) Y_{ij}^{(0)}$. 

Now we describe our model for the distribution of $U$, the treatment assignment, and the response. We use an independent Bernoulli random variable to model the residual confounding not captured by the observed covariates:
\[
U \mid \boldsymbol X \sim \text{Bernoulli}(p).
\] 
Here, we use a Bernoulli random variable to represent the unmeasured confounding for several reasons. First, it substantially reduces the computational cost of parameter estimation and is often adopted in the literature (\citealp{Rosenbaum1983}; \citealp{Imbens2003}; \citealp{Dorie2016}). Second, it facilitates contrasting our method to the bounds approach, where the unmeasured confounder takes on values $1$ and $0$ under the worst-case allocation. Third, while the support of $U$ (binary versus continuous) affects the interpretation of the association between $U$ and the treatment assignment and outcome, later we will reparametrize the association into a relative importance measure that is \emph{scale-free}. In Section \ref{sec: discussion} we will describe how our method could be extended to consider a continuous unmeasured confounder. 

We further assume that $(\boldsymbol{X},U)$ are all the confounders, observed and unobserved, so that conditional on $(\boldsymbol{X}, U)$, the treatment assignment is independent of potential outcomes, i.e., 
\[
Z_{ij} \indep (Y_{ij}^{(0)}, Y_{ij}^{(1)}) \mid \boldsymbol x_{ij}, u_{ij}. 
\]
We assume the treatment follows a logistic regression model:
\[
P(Z_{ij} = 1 \mid \boldsymbol x_{ij}, u_{ij}) = \text{expit}\{ \kappa(\boldsymbol x_{ij}) + \lambda u_{ij}\}  
\] and the response follows a normal model:\[
Y_{ij} \mid \boldsymbol x_{ij}, u_{ij}, z_{ij} \sim \text{Normal}(\psi(\boldsymbol x_{ij}) + \delta u_{ij} + \beta z_{ij}, \sigma^2). 
\label{eqn: dose/response assignment model}
\]
According to this formulation, $\beta$ denotes a constant and additive treatment effect, and $(\lambda, \delta)$ are sensitivity parameters that control how strongly the unobserved covariate $U$ is associated with the treatment assignment and the response. In our model, $p$ is also a sensitivity parameter. 

To prepare for calibration, we estimate not only the treatment effect, but also the effect of each observed covariate on treatment assignment and response, i.e., the coefficients of observed covariates in the outcome regression model and propensity score model. To estimate $\psi$ robustly, we leverage the matched design and write $\psi(\boldsymbol x_{ij}) = a_i + \boldsymbol\psi^T \boldsymbol x_{ij}$, where $\boldsymbol\psi^T \boldsymbol x_{ij}$ is a linear combination of $\boldsymbol x_{ij}$ and $a_i$ is a fixed effect specific to matched set $i$ (\citealp{Rubin1979}). Our model and inference are more robust compared to assuming $\psi(\boldsymbol x_{ij})$ is linear in $\boldsymbol x_{ij}$. For simplicity, we assume $\kappa$ is linear in $\boldsymbol x_{ij}$ here, but we may include a fixed effect for each matched set when estimating $\kappa(\cdot)$ as well.

The entire data-generating process is summarized below: 
\begin{equation}
    \begin{split}
        &U \mid \boldsymbol X = \boldsymbol x_{ij} \sim \text{Bern}(p) \\
        &P(Z_{ij} = 1 \mid \boldsymbol x_{ij}, u_{ij}) = \frac{\exp (\boldsymbol\kappa^T\boldsymbol x_{ij} + \lambda u_{ij}) }{1 + \exp(\boldsymbol\kappa^T\boldsymbol x_{ij} + \lambda u_{ij})} \\
        &Y_{ij} \mid \boldsymbol x_{ij}, u_{ij}, z_{ij} \sim \text{Normal}~(a_i + \boldsymbol\psi^T \boldsymbol x_{ij} + \delta u_{ij} + \beta z_{ij}, \sigma^2). 
    \end{split}
    \label{eqn: our model}
\end{equation}
To recap, we want to estimate $\boldsymbol\kappa$, $\boldsymbol\psi$, and the treatment effect $\beta$ for fixed sensitivity parameters $(p, \lambda, \delta)$, and compare the effect of this hypothesized hidden bias $U$ to the observed covariates in a meaningful way. In Supplementary Material C.1, we discuss how to extend the model to further accommodate treatment effect heterogeneity.

\section{Estimating Model Parameters via an EM Algorithm}
\label{sec: estimate param via EM}
We consider a similar model as in \citet{Rosenbaum1983} and \citet{Imbens2003}; however, we allow for fixed effects for matched sets. In \citet{Rosenbaum1983} and \citet{Imbens2003}'s models, there are only a handful of parameters, and these parameters can easily be estimated via maximum likelihood. In our model, to leverage the matched design, we allow for distinct fixed effects for each matched set and the number of parameters is of the same order as the number of matched sets. It is not unusual to have thousands of matched sets/pairs in a matched observational study and directly optimizing the observed data likelihood would be difficult. The key observation here is that the sensitivity analysis problem can be re-framed as a missing data problem (\citealp{Ichino2008}). Specifically, we treat the unmeasured confounder $U$ as a missing covariate and find the MLE efficiently via an Expectation Maximization (EM) algorithm, and then use the block bootstrap (\citealp{Abadie2017}) to construct the confidence interval. Below, we briefly describe the E-step and M-step in our context. A brief recap of the EM algorithm can be found in Supplementary Material B.1 and more details of the E-step and M-step can be found in Supplementary Material B.2.

Let $\boldsymbol x_\ell = (\boldsymbol x_{\text{obs},\ell}, \boldsymbol x_{\text{mis},\ell})$ denote the complete covariate data for each subject $\ell$, $\ell = 1,2, ..., N$, where $\boldsymbol x_{\text{mis},\ell} = u_{\ell}$ and $\boldsymbol x_\ell = (\boldsymbol x_{\text{obs},\ell}, u_\ell)$. Let $\boldsymbol{\gamma}$ denote the set of parameters of interest $(\boldsymbol\kappa, \boldsymbol\psi, \sigma, \beta)$, and $l(\boldsymbol\gamma\mid\boldsymbol x_\ell, z_\ell, y_\ell)$ the log-likelihood for the $\ell^{th}$ observation. The conditional expectation of the complete-data log-likelihood for subject $\ell$ is $Q_\ell = \sum_{u_\ell} l(\boldsymbol\gamma \mid \boldsymbol x_{\text{obs},\ell}, u_\ell, z_\ell, y_\ell)~p(u_\ell\mid\boldsymbol x_{\text{obs}, \ell}, y_\ell, z_\ell, \boldsymbol\gamma^{(s)})$, where $\boldsymbol\gamma^{(s)}$ denotes the current model parameters $(\boldsymbol\kappa^{(s)}, \boldsymbol\psi^{(s)},\sigma^{(s)}, \beta^{(s)})$ and the summation is over all possible realizations of the missing covariate $U$. For a fixed set of sensitivity parameters $(p, \lambda, \delta)$, the M-step reduces to finding $\boldsymbol\gamma = (\boldsymbol\kappa, \boldsymbol\psi,\sigma, \beta)$ that maximizes:
\begin{equation*}
\begin{split}
Q = \sum_{\ell = 1}^N Q_\ell &= \sum_{\ell = 1}^N \sum_{j = 0}^1 p(U_\ell = j\mid \boldsymbol x_{\text{obs},\ell}, y_\ell, z_\ell; \boldsymbol\gamma^{(s)})~l(\boldsymbol\gamma \mid \boldsymbol x_{\text{obs},\ell}, U_\ell = j, y_\ell, z_\ell) \\
&= \sum_{\ell = 1}^N \sum_{j = 0}^1 w_{\ell j}\cdot \big\{ l_{y_\ell \mid \boldsymbol x_\ell, z_\ell}(\boldsymbol\psi, \sigma, \beta) + l_{z_\ell\mid\boldsymbol x_\ell}(\boldsymbol\kappa)\big\}\\
&= \sum_{\ell = 1}^N \sum_{j = 0}^1 w_{\ell j} ~l_{y_\ell \mid \boldsymbol x_\ell, z_\ell}(\boldsymbol\psi, \sigma, \beta) + \sum_{\ell = 1}^N \sum_{j = 0}^1 w_{\ell j} ~l_{z_\ell \mid \boldsymbol x_\ell}(\boldsymbol\kappa)\\
&= Q^{(1)} + Q^{(2)},
\end{split}
\end{equation*}
where \[
w_{\ell j} = P(U_\ell = j\mid \boldsymbol x_{\text{obs},\ell}, y_\ell, z_\ell; \boldsymbol\gamma^{(s)}).
\]
To maximize $Q$, it suffices to maximize $Q^{(1)}$ and $Q^{(2)}$ separately, which reduces to finding the MLE for a weighted regression and a weighted logistic regression, and can be easily implemented in commonly used statistical software.

\section{Calibrating the Unmeasured Confounder to Observed Covariates}
\label{sec: calibration}
The parameters in the bounds approach are interpreted in terms of the odds ratio. For instance, in \citet{Gastwirth1998}'s simultaneous sensitivity analysis, two subjects matched for observed covariates $\boldsymbol x$ differ in their odds of receiving treatment by at most a factor of $\exp(\lambda)$. The magnitude of the sensitivity parameter $\lambda$ speaks of the strength of the hypothesized unmeasured confounder in confounding the treatment assignment; however, an audience without much training in interpreting sensitivity analysis results may still be perplexed about what $\lambda$ means for their specific problem. Calibration aims to remedy this by comparing the hypothesized unmeasured confounder to the observed covariates in a meaningful way. We ask two interconnected questions. First, had there existed a binary unmeasured confounder, how big an impact, compared to the observed covariates, would it have to have on the treatment assignment and the response in order to materially affect the qualitative conclusion we draw from the observational data? Second, what can we say about the importance of this unmeasured confounder relative to that of the other observed covariates? 

We leverage our proposed sensitivity analysis model to answer these two questions. For a specified parameter $p$, we first identify the boundary that separates $(\lambda, \delta)$ pairs that render the treatment effect significant at $0.05$ level from those that do not. In practice, for each fixed $\lambda$, we do a binary search to find the largest $\delta$ that renders the treatment effect significant. For a specific $(\lambda, \delta)$ pair, we record the coefficient estimates of observed covariates and contrast them with $(\lambda, \delta)$ on the same graph. To make the comparison between the coefficient estimates and the sensitivity parameters meaningful, we transform each observed covariate to the same scale as the hypothesized unmeasured confounder. Here, we follow the suggestion in \citet{Gelman2008} and standardize all non-binary covariates to mean $0$ and SD $0.5$ and leave dichotomized covariates intact. According to this standardization scheme, the coefficients of dichotomized variables can be interpreted directly and those of continuous/ordinal variables can be interpreted as the effect of a 2-SD increase in the covariate value, which roughly corresponds to flipping a binary variable from $0$ to $1$. This calibration plot enables empirical researchers to better interpret the sensitivity parameters in a matched observational study by helping them draw comparisons to the more tangible observed covariates. When we implement the calibration method in our R package \textsf{sensitivityCalibration}, we make an interactive calibration plot where for a fixed $p$, the $(\lambda, \delta)$ pair moves along the boundary and the coefficients of the observed covariates adjust themselves accordingly.

Comparing the sensitivity parameters to the coefficients of observed covariates sheds some light on interpreting the magnitude of the sensitivity parameters; however, a naive comparison of sensitivity parameters to regression coefficients may not suffice for two reasons. First, although $U$ is assumed to be independent of all observed covariates $\mathbf{X}$, any two observed observed covariates $X_i$ and $X_j$ may be correlated, and the magnitudes of their regression coefficients $|\beta_i|$ and $|\beta_j|$ are not necessarily informative about how important $X_i$ and $X_j$ are in confounding the treatment and outcome. Second, sometimes researchers would like to calibrate the unmeasured confounder to a group of two or more observed covariates, say ethnicity and poverty income ratio. It is not clear how to achieve this by simply focusing on regression coefficients' magnitudes.

To tackle these challenges, we propose to supplement this naive comparison with an assessment of the \emph{relative importance} of the hypothesized unmeasured confounder compared to the observed covariates. Many relative importance measures have been developed over the years; see \citet{Kruskal1989} for a taxonomy of these measures. A ``well-known, venerable collection'' of measures (\citealp{Kruskal1989}) often express the relative importance of covariates in terms of the variance reduction, i.e., how much variance of the response is accounted for by each covariate. Specifically, in the context of multiple regression, \citet{Pratt1987} derived a unique measure of variable importance based on symmetry and invariance principles under linear transformation. This measure assigns to each variable an importance that has the interpretation of an average reduction in variance if that variable is held fixed. More recently, \citet{Azen2003} proposed the so-called dominance analysis, a method based on examining R-squared values across all subset models.

Our sensitivity analysis framework works seamlessly with such relative importance measures and enables researchers to speak of the importance of hypothesized unmeasured confounder relative to the observed covariates. For each sensitivity parameter trio $(p, \lambda, \delta)$, we run the EM algorithm until it converges, and at convergence we have a dataset where each original instance is augmented into two instances, one with the binary unmeasured confounder $U = 0$ and the other with $U = 1$. Each of them is associated with a weight, which is the probability that $U = 1$ or $U = 0$ given the data and parameter values at convergence. We expand this weighted dataset and obtain a full dataset that contains all the observed covariates as well as the binary unmeasured confounder $U$, and use this full dataset to assess the relative importance of covariates. It is also straightforward to assess the relative importance of a group of covariates under our framework: Pratt's method has the property that the relative importance of several variables is the sum of their individual relative importance. On the other hand, dominance analysis handles a group of covariates by adding the entire group into each subset model and computing the average increase of $R^2$. Note that regression coefficients of observed covariates, as well as their relative importance measures, are all computed with adjustment for the unmeasured confounder. As pointed out in Section \ref{subsec: contribution}, this is an important distinction between our proposal and many existing calibration approaches in the literature. We leverage proposed calibration procedures to study in detail the causal relationship between second-hand smoke exposure and blood lead levels in Section \ref{subsec: calibration}.

\section{Simulation Study}
\label{sec: simulation}
In a simulation study, we examined the confidence interval coverage of our approach and how our approach
compares in power to the bounds approach under the simultaneous sensitivity model of \citet{Gastwirth1998}. The power of a sensitivity analysis is defined as the probability that it correctly rejects the null hypothesis in a favorable situation where there is a genuine treatment effect and no unmeasured confounding (\citealp{Rosenbaum2010design_sens}). If there is no bias and a genuine treatment effect, then we would hope to reject the null hypothesis of no effect and the power of a
sensitivity analysis is the chance that our hope will be realized. The power of a sensitivity analysis guides the choice of methods of analysis for a fixed research design (\citealp{hansen2014clustered}).

In Subsection \ref{subsec: simulation: linear}, we generate data according to the data generating process described in Section \ref{section: notation and model} with $\kappa(\boldsymbol x_i)$ and $\psi(\boldsymbol x_i)$ both linear in $\boldsymbol x_i$, a real treatment effect, and no unmeasured confounding (corresponding to $(\lambda, \delta) = (0, 0)$). We verify that our confidence interval has correct coverage and demonstrate that our method is more powerful than the simultaneous sensitivity analysis in \citet{Gastwirth1998}. In Subsection \ref{subsec: simulation: non-linear}, we generate the response from a nonlinear model and verify that matching combined with regression adjustment within matched sets makes our method robust against model misspecification. In Subsection \ref{subsec: simulation: more}, we perform additional simulations with different effect sizes, error structures, and consider more realistic situations where the matching is not exact. In all of these scenarios, we demonstrate that our framework is robust against various model misspecifications, including a non-normal error structure, a non-linear response model, and non-exact matching, in the sense that the coverage of our confidence intervals is always approximately equal to the nominal coverage. 

\subsection{Linear response model}
\label{subsec: simulation: linear}
We generate $100$ strata with $10$ subjects in each stratum. Subjects in the same stratum have the same observed covariates $\boldsymbol x_{obs}$, which consists of $7$ measurements, each independently and normally distributed with mean $(3, 1, 5, 2, 6, 4, 5)$ and standard deviation $(1, 0.15, 1.5, 0.2, 1, 0.8, 1)$. Each subject is further associated with an unmeasured confounder $U$ such that $U \mid \boldsymbol X \sim \text{Bern}(0.5)$. We simulate a favorable situation where there is a real treatment effect and no unmeasured confounding by letting $\beta = 2$, $\sigma = 1.5$, and $\lambda = \delta = 0$, and setting both the treatment assignment and response model linear in $\boldsymbol x_{obs}$ as follows:
\begin{equation*}\small
    \begin{split}
        &P(Z = 1 \mid \boldsymbol X, U) = \text{expit}(-0.03X_1 + 0.08X_2 + 0.02X_3 -0.9X_4 + 0.6X_5 -0.5X_6 + 0.7X_7 - 1.5), \\
&Y \mid X, U, Z = 0.1X_1 - 0.08X_2 + 0.04X_3 -0.9X_4 + 2X_5 -0.5X_6 + X_7 - 5 + 2Z + \epsilon, \\
&\epsilon \sim N(0, 1.5^2).
    \end{split}
\end{equation*}
We estimate $\beta$ from the simulated dataset $(\boldsymbol x_{obs}, Z, Y)$ for various sensitivity parameters. We repeat the simulation $500$ times and use $500$ bootstrap samples to construct the confidence interval each time. With $100$ strata and $10$ subjects in each stratum, we observed that $94.4\%$ of the times the $95\%$ confidence intervals cover the true parameter $\beta = 2$ when $(\lambda, \delta)$ are set at the truth of $0$. We also compare the power of our sensitivity analysis to the simultaneous sensitivity analysis by \citet{Gastwirth1998}. Table \ref{tbl: our model power linear and nonlinear} summarizes the power of both approaches for various $(\lambda, \delta)$ values in this case. Our approach is much more powerful: when $(\lambda, \delta) = (2.5, 2.5)$, our approach correctly rejects the null hypothesis $95.5\%$ of the times, while the bounds approach for the simultaneous sensitivity analysis only correctly rejects the null hypothesis $19.8\%$ of the times. We repeat the simulation with a larger sample size ($200$ strata with $20$ subjects in each stratum) and the same qualitative results hold. Next, we replace the constant additive treatment effect $\beta = 2$ with a heterogeneous treatment effect $\beta \sim N(2, 1)$. We compute the level of our approach by calculating how many $95\%$ CIs capture $\mathbb{E}[\beta_i] = 2$, and power by calculating how many $95\%$ CIs do not contain $0$. Table \ref{tbl: our model power linear and nonlinear} summarizes the results in this scenario and again similar qualitative results hold.

\begin{table}
\caption{\label{tbl: our model power linear and nonlinear}\small
A comparison of power of sensitivity analysis. From left to right: linear model with a constant additive treatment effect and sample size $1000$; linear model with a constant additive treatment effect and sample size $4000$; linear model with an additive but heterogeneous treatment effect and sample size $1000$; non-linear model with a constant additive treatment effect and sample size $1000$. Levels of our method are $5.6\%$, $4.5\%$, $4.2\%$, and $5.4\%$, respectively, when $(\lambda, \delta) = (0, 0)$ are set as the truth and the nominal level is $5\%$.}
\centering
\resizebox{\columnwidth}{!}{
\fbox{%
\begin{tabular}{ccccccccc}
  \hline
  &\multicolumn{2}{c}{Linear, $n = 1000$} &\multicolumn{2}{c}{Linear, $n = 4000$} &\multicolumn{2}{c}{Heterogeneous, $n = 1000$}&\multicolumn{2}{c}{Non-linear, $n = 1000$} \\ 
  \hline
\multirow{3}{*}{\begin{tabular}{c} $(\lambda, \delta)$ \end{tabular}} & \multirow{3}{*}{\begin{tabular}{c} Full matching\\ with regression\\ adjustment \end{tabular}} 
& \multirow{3}{*}{\begin{tabular}{c} Simultaneous\\ sensitivity\\ analysis \end{tabular}} & \multirow{3}{*}{\begin{tabular}{c} Full matching\\ with regression\\ adjustment \end{tabular}} 
& \multirow{3}{*}{\begin{tabular}{c} Simultaneous\\ sensitivity\\ analysis \end{tabular}} 
& \multirow{3}{*}{\begin{tabular}{c} Full matching\\ with regression\\ adjustment \end{tabular}} 
& \multirow{3}{*}{\begin{tabular}{c} Simultaneous\\ sensitivity\\ analysis \end{tabular}}
& \multirow{3}{*}{\begin{tabular}{c} Full matching\\ with regression\\ adjustment \end{tabular}} 
& \multirow{3}{*}{\begin{tabular}{c} Simultaneous\\ sensitivity\\ analysis \end{tabular}} \\ \\ \\
  \hline
(0, 0) & 1.000 & 1.000 & 1.000 & 1.000 &1.000 &1.000 &1.000 &1.000\\ 
(1, 1) & 1.000 & 1.000 & 1.000 & 1.000 &1.000 &1.000 &1.000 &1.000\\ 
(1.5. 1.5) & 1.000 & 1.000 & 1.000 & 1.000 &1.000 &1.000 &1.000 &1.000\\ 
(2, 2) & 1.000 & 0.998 & 1.000 & 1.000 &1.000 &0.840 &1.000 &0.824\\ 
(2.5, 2.5) & 0.954 & 0.198 & 1.000 & 0.672 &0.952 &0.012 &0.928 &0.076\\ 
(3, 3) & 0.000 & 0.000 & 0.000 & 0.000 &0.200 &0.000 &0.000 &0.000\\ 
   \hline
\end{tabular}}}
\end{table}

\subsection{Nonlinear response model}
\label{subsec: simulation: non-linear}
One advantage of matching is that it renders the model more robust against misspecification. To demonstrate this, we simulate datasets according to the same propensity score model but a highly non-linear response model as follows: 
\begin{equation*}
    \begin{split}
        &P(Z = 1 | \boldsymbol X, U) = \text{expit}(-0.03X_1 + 0.08X_2 + 0.02X_3 -0.9X_4 + 0.6X_5 -0.5X_6 + 0.7X_7 - 1.5), \\
&Y \mid X, U, Z = 0.7X_1^2 - 0.8X_2^3 - 0.7|X_3|^{1/3}+ 0.2X_5^2 - X_6 + 2X_7 + 2Z + \epsilon, ~\epsilon \sim N(0, 1.5^2).
    \end{split}
\end{equation*}

It would be difficult, if not impossible, to correctly specify this response model; however, as one performs matching and performs regression adjustment within each matched set, linear regression with matched-set fixed effects works well in approximating the true response surface. In fact, $94.6\%$ of our $500$ experiments have $95\%$ confidence intervals covering the true treatment effect $\beta = 2$. Moreover, a comparison of the power of sensitivity analysis again demonstrates that our model has superior power compared to the simultaneous sensitivity analysis, as shown in Table \ref{tbl: our model power linear and nonlinear}. 

\subsection{Different effect sizes, error structures, and non-exact matching}
\label{subsec: simulation: more}
In this subsection, we report additional simulation results for different effect sizes, error structures, and non-exact matching. We first consider the same linear and non-linear response model as in subsection \ref{subsec: simulation: linear} and \ref{subsec: simulation: non-linear} but with a smaller effect size $\beta = 1$. With the linear response model, $93.2\%$ of confidence intervals cover the true parameter $\beta = 1$ in our simulations, and the confidence intervals have almost exact coverage for the non-linear response model, with $94\%$ of the $95\%$ confidence intervals cover the true parameter. Table \ref{tbl: our model power effect size = 2/3 and non-normal error} further summarizes the power of our sensitivity analysis in this setting.

\begin{table}
\caption{\label{tbl: our model power effect size = 2/3 and non-normal error}\small Power of sensitivity analysis. Column 2 and 3: linear and non-linear model with effect size = $2/3$ and sample size $1000$. Levels are $6.0\%$ and $5.0\%$, respectively, when $(\lambda, \delta) = (0, 0)$ are set at truth. Column 4 and 5: linear response model with non-normal error. Levels are $4.6\%$ and $6.2\%$ when the error is distributed as a double exponential distribution and a Student's t distribution (df = 2), respectively, when $(\lambda, \delta) = (0, 0)$ are set as the truth and the nominal level is $0.05$.
}
\centering
\fbox{%
\begin{tabular}{ccccccc}
  \hline
  \hline
  &\multicolumn{2}{c}{Effect size = 2/3} &\multicolumn{2}{c}{Non-normal error}\\ \hline
$(\lambda, \delta)$ & Linear response &Non-linear response & Double exponential & Student's t (df = 2)\\
  \hline
(0, 0) & 1.000 &  1.000 &1.000  & 0.960\\ 
(0.5, 0.5) & 1.000 & 1.000 &1.000 & 0.954\\ 
(0.8. 0.8) & 1.000 & 1.000 &1.000 &0.932\\ 
(1, 1) & 1.000 & 1.000 &1.000 & 0.914\\ 
(1.5, 1.5) & 0.954 & 0.964 &0.944 & 0.626\\ 
(2, 2) & 0.008 & 0.014 &0.008  & 0.028\\ 
   \hline
\end{tabular}}
\end{table}
\thispagestyle{empty}

Next, instead of assuming a normal error, we consider two different error structures: a Student's t distribution with $2$ degrees of freedom and a Laplace (double exponential) distribution with rate $1.5$. We adopt a linear response model with an additive treatment effect $\beta = 1$, no unmeasured confounding ($\lambda = \delta = 0$), and sample size equal to $1000$. $95.4\%$ and $93.8\%$ of confidence intervals capture the truth $\beta = 1$ when $\epsilon$ takes on a Student's t distribution with $2$ degrees of freedom and a Laplace distribution with rate $1.5$ respectively. Again, the results demonstrate that our method is robust against misspecified error structures. The power of our sensitivity analysis is further summarized in Table \ref{tbl: our model power effect size = 2/3 and non-normal error}. 

In all aforementioned simulations, we made it easy on matching by simulating multiple objects with exactly the same observed covariates, which is rarely the case in practice. To better mimic a real dataset, we add a uniform $[-0.2, 0.2]$, uniform $[-0.5, 0.5]$, or uniform $[-1.0, 1.0]$ noise to each observed covariate so that subjects in the same matched set are similar, but no longer identical in their observed covariates. We run the experiment with the same linear response model as specified before with $\beta = 1$ and $\sigma = 1.5$. When we perform the matching, subjects with different observed covariates, sometimes from different strata, are put in the same matched set. As a consequence, hypothesis testing assuming a homogeneous, additive treatment effect without adjustment for covariates may yield invalid inferences. However, our method adjusts for covariates and produces valid inferences if the model for covariates is correctly specified.  Our confidence intervals covered the true parameter $92.4\%$, $95.6\%$ and $94.4\%$ of the times for the uniform $[-0.2,.02]$, $[-0.5,0.5]$, and $[-1,1]$ noise, respectively. Table \ref{tbl: our model power not exact match linear} summarizes the power of our sensitivity analysis. Our method is still quite powerful: it correctly rejects the null hypothesis in a favorable situation $94.4\%$ of the times when $(\lambda, \delta) = (1.5, 1.5)$, while the simultaneous sensitivity analysis only rejects $62.6\%$ of the times. 

\begin{table}
\caption{\label{tbl: our model power not exact match linear}
\small Power of sensitivity analysis: linear response model with non-exact matching. Levels are $7.6\%$, $4.4\%$, and $5.6\%$ for different degrees of inexact matching, when $(\lambda, \delta) = (0, 0)$ are set as the truth and nominal level is $5\%$.}
\centering
\fbox{%
\begin{tabular}{rrrr}
  \hline
  \hline
$(\lambda, \delta)$ & Uniform [-0.2, 0.2] &Uniform [-0.5, 0.5] & Uniform [-1, 1]\\ 
  \hline
(0, 0) &1.000  &1.000 & 1.000\\ 
(0.5, 0.5) &1.000 &1.000 & 1.000\\ 
(0.8. 0.8) &1.000 &1.000 & 1.000\\ 
(1, 1) &1.000 &1.000 & 1.000\\ 
(1.5, 1.5) &0.944 &0.878 & 0.772\\ 
(2, 2) &0.008  &0.008 & 0.008 \\ 
   \hline
\end{tabular}}
\end{table}
\pagestyle{empty}

\section{Application: Second-Hand Smoking and Blood Lead Levels in Children}
\label{sec: case study}
\subsection{Robustness of the causal relationship between second-hand smoking and blood lead levels in children}
We now leverage our proposed sensitivity analysis method to examine how sensitive the treatment effect is to unmeasured confounding in the second-hand smoke exposure and blood lead levels study described in Section \ref{section: intro}. For a fixed sensitivity parameter trio $(p, \lambda, \delta)$, we maximized the observed data likelihood of model \eqref{eqn: our model} as described in Section \ref{sec: estimate param via EM} after conducting a full match. Model \eqref{eqn: our model} assumes that the error term in the blood lead level model is normally distributed. Simulations in Section \ref{subsec: simulation: more} suggest that matching combined with regression adjustment is at least somewhat robust to non-normal errors. We first tabulate the $95\%$ confidence intervals of treatment effect for various combinations of sensitivity parameters $(p, \lambda, \delta)$ in Table \ref{tbl: our_result}. Additional results for more $(p, \lambda, \delta)$ combinations are presented in Supplementary Materials C.2. Note that $(\lambda, \delta) = (0, 0)$ corresponds to no unmeasured confounding and we would expect the same confidence interval regardless of $p$. This is indeed the case. We compare our method to the bounds approach of \citet{Gastwirth1998}. The last column in Table \ref{tbl: our_result} summarizes the worst-case p-values of testing no treatment effect with various $(\lambda, \delta)$ pairs. When $(\lambda, \delta) = (1.0, 1.0)$, the $95\%$ confidence interval does not cover $0$ for $p = 0.1$, $p = 0.3$, or $p = 0.5$; however, the bounds approach of \citet{Gastwirth1998} yields a p-value of $0.630$. The null hypothesis cannot be rejected according to the bounds approach; however, an unmeasured confounder of such magnitude cannot explain away the treatment effect under our sensitivity analysis model.

\begin{table}
\caption{\label{tbl: our_result} Treatment effect versus different (p, $\lambda$, $\delta$) pair}
\centering
\fbox{%
\begin{tabular}{ccccc}
\hline
                & p = 0.5          & p = 0.3    & p = 0.1  & \multicolumn{1}{c}{\text{Simultaneous}}\\
\hline
($\lambda$, $\delta$) & 95\% CI    & 95\% CI    & 95\% CI   & \multicolumn{1}{c}{p-value}\\ \hline
(0,0)           & (0.397, 0.808)   & (0.407, 0.823) &(0.415, 0.825) & 0\\
(0.5, 0.5)      & (0.349, 0.750)   & (0.345, 0.766) &(0.380, 0.771) & 7.55e-15\\
(0.8, 0.8)      & (0.243, 0.673)   & (0.275, 0.677) &(0.341, 0.749) & 6.93e-4\\
(1.0, 1.0)      & (0.144, 0.572)   & (0.180, 0.594) &(0.317, 0.711) & 0.488 \\
(1.2, 1.2)      & (0.0448, 0.455)  & (0.106, 0.507) &(0.259, 0.676) & 0.999 \\ 
(1.5, 1.5)      & (-0.142, 0.276) & (-0.0670, 0.323) &(0.212, 0.602)& 1 \\
(2.0, 2.0)      &(-0.551, -0.141)  &(-0.442, -0.0501) &(0.0491, 0.474) & 1 \\
\hline
\end{tabular}}
\end{table}

\begin{figure}
\subfloat[\small $(\lambda, \delta) = (1, 1)$]{\includegraphics[width = 2.7in]{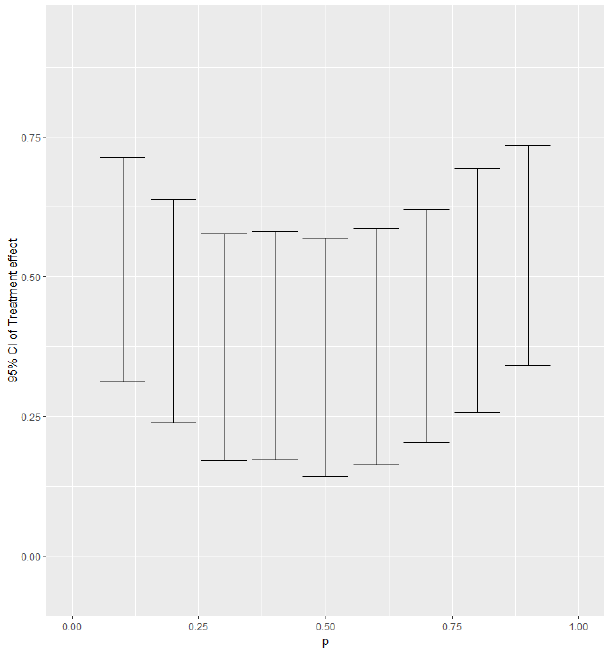}} \hspace{0.2 cm}
\subfloat[\small$(\lambda, \delta) = (1.5, 1.5)$]{\includegraphics[width = 2.7in]{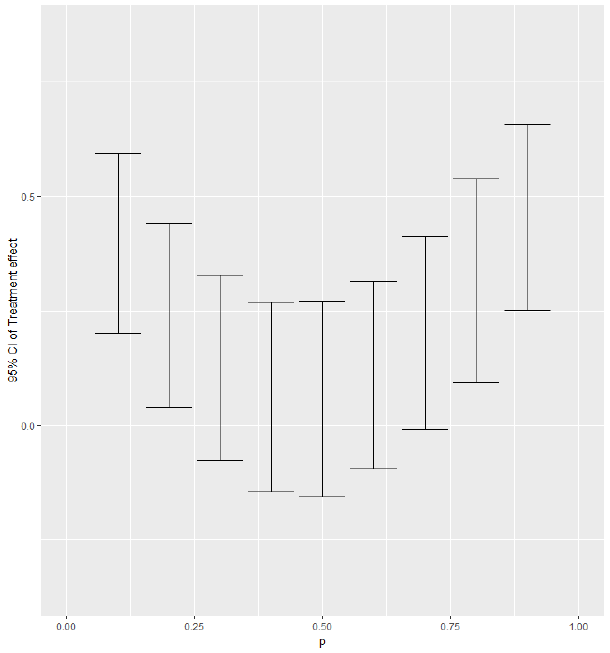}}
 \caption[$95\%$ CI versus $p$ for different $(\lambda, \delta)$ pairs]
        {\small $95\%$ CI versus $p$ for different $(\lambda, \delta)$ pairs} 
        \label{fig: CI versus p: different lambda and delta}
\end{figure}

For a fixed $(\lambda, \delta)$ pair, a public health researcher may be interested in plotting the $95\%$ confidence interval against all values of $p$ between $0$ and $1$ and reporting the inference for different $p$. See Figure \ref{fig: CI versus p: different lambda and delta} for such a plot for various $(\lambda, \delta)$ combinations. Figure \ref{fig: CI versus p: different lambda and delta} suggests that the choice of $p = 0.5$ yields approximately the most conservative treatment effect for this study. Intuitively, when $p$ is close to $0$ or $1$, subjects with different treatment status in the same matched set tend to have the same value of $U$, which makes it difficult to attribute the observed treatment effect to the disagreement in $U$; therefore, we would expect that $p$ taking a value somewhere in between either ends would yield the most conservative inference. Instead of reporting Figure \ref{fig: CI versus p: different lambda and delta}, empirical researchers may simply report a summary statistic. For a fixed $(\lambda, \delta)$ pair, let $\text{CI}_p = [L_p, U_p]$ represent the level $\alpha$ CI corresponding to $p$. One choice is to report the most conservative $\text{CI}_p$, i.e., $\text{CI}_{p^\ast}$, where $p^\ast = \text{argmin}_{p \in [0, 1]} ~L_p$. One may also report $\cup_{p \in [0, 1]} \text{CI}_p$ so that the confidence interval has the correct (but conservative) coverage for all realizations of $p$.

A public health researcher studying this particular dataset could stop here and report the sensitivity analysis results as discussed above. However, the subject matter expert, as well as any audience of the report, may stare at the sensitivity analysis, rather confused. The absolute scale of $(\lambda, \delta)$ sheds little light on the actual interpretation of this hypothesized unmeasured confounder: the causal conclusion is still significant when $(\lambda, \delta) = (1.2,1.2)$ for $p = 0.1, 0.3$ and $0.5$, but no longer so when $(\lambda, \delta) = (1.5, 1.5)$. Is the hypothesized unmeasured confounding indeed something worth worrying about? Does this degree of sensitivity to unmeasured confounding cast doubt on the study's findings when no unmeasured confounding was assumed or does it reinforce the findings? To help subject matter experts better interpret the sensitivity parameters in their specific context, we compare the hypothesized unmeasured confounder to the observed covariates below.

\subsection{Calibration to observed covariates}
\label{subsec: calibration}
\begin{figure}[h]
\includegraphics[width = \textwidth]{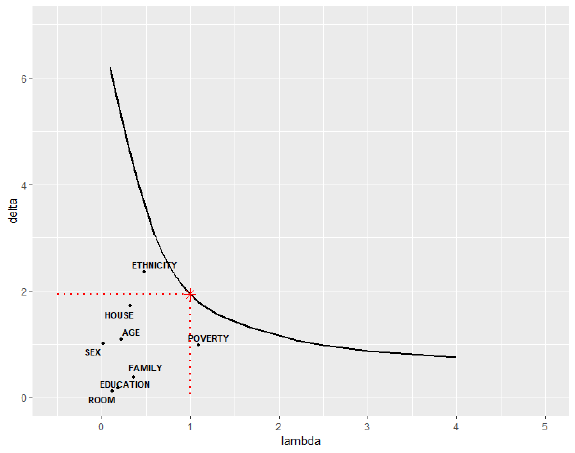}
\caption{\small A calibration plot with $(p, \lambda, \delta) = (0.5, 1, 1.99)$. Labels and the covariate information they encode are as follows: \textbf{POVERTY} (poverty income ratio), \textbf{AGE} (age at the time of interview), \textbf{SEX} (male/female), \textbf{HOUSE} (whether the house is built before or after 1974), \textbf{ROOM} (number of rooms in the house), \textbf{FAMILY} (size of family), \textbf{ETHNICITY} (white/non-white), \textbf{EDUCATION} (years of education of the reference adult).}
\label{fig: calibrate plot zoom}
\end{figure}

Figure \ref{fig: calibrate plot zoom} displays our calibration plot when $p = 0.5$. The marker represents one $(\lambda, \delta)$ pair on the boundary and for this specific sensitivity parameter pair $(\lambda, \delta) = (1, 1.99)$, we estimate the coefficients of observed covariates (after standardized to mean $0$ and SD $0.5$ if necessary), which are shown on the same plot as dots with labels. A subject matter expert can make sense of Figure \ref{fig: calibrate plot zoom} as follows: in order for the binary unmeasured confounder $U$ to explain away the treatment effect, it may be associated with sensitivity parameters $(\lambda, \delta) = (1, 1.99)$. A $(\lambda, \delta)$ pair as large as $(1, 1.99)$ says that flipping the unmeasured confounder from $0$ to $1$ corresponds to a one-unit increase in the propensity score (in logit scale) and almost a 2 $\mu$g/dL increase in the blood lead level, holding everything else fixed. To draw a comparison, we see a 2-SD increase in the poverty-to-income ratio roughly corresponds to a 1.2-unit increase in the propensity score and a 1 $\mu$g/dL increase in the blood lead level, while holding everything else fixed. We implement an interactive calibration plot in our R package \textsf{sensitivityCalibration} and Supplementary Materials C.3 contains $4$ snapshots of this animated plot with different $(\lambda, \delta)$ pairs on the boundary. The estimated coefficients of the observed covariates are relatively insensitive to the change in $(\lambda, \delta)$.

As is discussed in Section \ref{sec: calibration}, magnitudes of regression coefficients, even after standardization, do not necessarily speak of their importance. We adopt two useful and interpretable metrics to describe the importance of the hypothesized unmeasured confounder relative to that of other observed covariates. In the outcome regression model, we assess the relative variable importance using both Pratt's method (\citealp{Pratt1987}) and the dominance analysis method (\citealp{Azen2003}). The results are summarized in Table \ref{tbl: relative importance OR}. In Pratt's method, the contribution of each covariate is the share of total $R^2$ this covariate accounts for, while in the dominance analysis, the contribution of each covariate is the average increase in $R^2$ across all subset models. Although the two analyses have different motivations, the results align perfectly. With $\delta = 2$, the hypothesized binary unmeasured confounder $U$ is almost comparable to the observed covariate POVERTY (the poverty income ratio) in confounding the blood lead level. Similarly, we assess the relative importance of $U$ in confounding the propensity score under the logistic regression model via the generalized dominance analysis. With $\lambda = 1$, the hypothesized unmeasured confounder in fact has higher relative importance than any other observed covariates in confounding the treatment assignment. This supplementary information helps subject matter experts judge for themselves how robust the causal conclusion is.  

\begin{table}
\caption{\label{tbl: relative importance OR}
\small Assessing the relative importance of observed confounders and the hypothesized unmeasured confounder:  \textbf{POVERTY} (poverty income ratio), \textbf{AGE} (age at the time of interview), \textbf{SEX} (male/female), \textbf{HOUSE} (whether the house is built before or after 1974), \textbf{ROOM} (number of rooms in the house), \textbf{FAMILY} (size of family), \textbf{ETHNICITY} (white/non-white), \textbf{EDUCATION} (years of education of the reference adult), \textbf{COP} (treatment status).}
 \centering
 \fbox{%
\begin{tabular}{ccccccc}
  \hline
  \hline
  \multicolumn{4}{c}{Outcome Regression ($\delta = 2$)} & \multicolumn{2}{c}{Propensity score ($\lambda = 1$)}\\
  \multicolumn{2}{c}{Pratt's method} &\multicolumn{2}{c}{Dominance analysis} &\multicolumn{2}{c}{Generalized dominance analysis} \\ 
  \hline
Covariate & Contribution & Covariate & Contribution     & Covariate & Contribution\\  
  \hline
AGE         & 0.044   & AGE         & 0.043           &\textbf{U} & 0.045  \\ 
POVERTY   	& 0.040   & POVERTY     & 0.035           &POVERTY   & 0.024 \\  
\textbf{U} 	& 0.032   & \textbf{U}   & 0.031           &HOUSE     & 0.005\\ 
ETHNICITY 	& 0.023   & ETHNICITY    & 0.020           &ETHNICITY   & 0.004\\ 
FAMILY     	& 0.018   & FAMILY 	& 0.019           & FAMILY      & 0.003 \\ 
HOUSE 	    & 0.018   & HOUSE      & 0.017           & ROOM      & 0.003\\ 
SEX         & 0.013   & SEX        & 0.013           & AGE   & 0.003\\ 
COP         & 0.010   & COP       & 0.012           & EDUCATION  & 0.001\\ 
EDUCATION   & 0.007  & EDUCATION  & 0.011          & SEX  & 0.000\\ 
ROOM        & -0.003 & ROOM       & 0.002 \\
\hline
\end{tabular}}
\end{table}

We are now able to make causal statements regarding the detrimental effect of second-hand smoke on a child's blood lead level as follows. Under the assumption of treatment ignorability, the exposure to second-hand smoke causes a significant increase in blood lead levels in children, with a $95\%$ confidence interval being $(0.397, 0.808)$ $\mu$g/dL. To explain away this causal conclusion, the residual confounding represented by an independent binary unmeasured confounder needs to be associated with the blood lead level to a similar extent as the poverty income ratio and at the same time be more associated with the cotinine level than any of the eight observed covariates.

\section{Discussion}
\label{sec: discussion}
In this paper, we investigate the detrimental effect of second-hand smoking on blood lead levels among children using observational data. To better assess the robustness of our causal conclusion, we develop a new sensitivity analysis framework for matched observational studies that allows calibration, i.e., the hypothesized unmeasured confounder can be meaningfully compared to the observed covariates. Our sensitivity analysis has two outputs. The first output is a calibration plot. For a fixed sensitivity parameter $p$, we identify $(\lambda, \delta)$ pairs such that the estimated treatment effect is significant at the $0.05$ level, and for each $(\lambda, \delta)$ pair on the boundary, we further estimate the coefficients of observed covariates by doing regression adjustment in each matched set, and contrast them with $(\lambda, \delta)$ on the same plot. To further complement a naive comparison of magnitudes of regression coefficients to sensitivity parameters, we supplement the calibration plot with a table summarizing the variable importance of the unmeasured confounder relative to other observed covariates using both Pratt's method and the dominance analysis. This allows us to further assess the robustness of our causal conclusions regarding the detrimental effect of second-hand smoke on blood lead levels. We find that in order to explain away the observed causal relationship, the unmeasured confounding needs to be associated with blood lead levels to a similar degree as the poverty income ratio, and more associated with cotinine level than any observed covariate.

In this paper, we model the unmeasured confounder as binary. We discussed motivations behind this modeling choice in Section 2. This modeling aspect can be generalized and is not essential in principle. When the unmeasured confounder has a continuous distribution, the E-step yields an integral rather than a weighted sum and this integral typically has no closed form; however, it can be approximated by being discretized into a weighted sum so that the EM algorithm by the method of weights can still be applied (\citealp{Brahim1992}). Alternatively, one may use a Monte Carlo EM algorithm (\citealp{Ibrahim1990}). These approaches usually involve heavy computation. Hence, we focus on analyzing a binary unmeasured confounder in the paper, although the extension to a continuous unmeasured confounder with any pre-specified distribution is immediate. For future work, it would be interesting to investigate the consequences of assuming a dichotomous unmeasured confounder. In the context of hypothesis testing, it is shown in \citet{Wang2006} that the causal conclusions are most sensitive to a binary unobserved covariate for a large class of test statistics in the setting of matched pairs with binary treatment and response. Any analogous result in the context of estimation would be of great interest. 

\bibliographystyle{rss} 
\bibliography{calibration_1}

\clearpage
\pagenumbering{arabic}

  \begin{center}
    {\LARGE\bf Supplementary Materials for ``A Calibrated Sensitivity Analysis for 
Matched Observational Studies 
with Application to the Effect of Second-hand 
Smoke Exposure on Blood Lead Levels in Children"}
\end{center}
  \medskip

\begin{center}
{\large\bf Supplementary Material A: Literature Review}
\end{center}

\section*{A.1: Model-based approaches to sensitivity analysis}
The Rosenbaum bounds approach considers a family of possible distributions for the unmeasured confounder which is restricted by sensitivity parameters and then finds the worst-case distribution (with respect to finding evidence for a treatment effect) for the unmeasured confounder in this family. \citet{Rosenbaum1983} and \citet{Imbens2003} took a different model-based approach that fully specifies the distribution of the unmeasured confounder. Under the setting of a binary response and binary treatment, \citet{Rosenbaum1983} assumed a discrete stratification variable $\boldsymbol S$ and within each stratum, they specified the distribution of $U \sim \text{Bern}(p)$ with a sensitivity parameter $p$, in addition to one sensitivity parameter that controls the effect of $U$ on the odds of treatment, one that controls the effect of $U$ on the odds of the outcome when treated, and one that controls the effect of $U$ on the odds of the outcome when not treated. Given these four sensitivity parameters, the effect of the treatment on the odds of the outcome is identified and can be estimated via maximum likelihood. \citet{Imbens2003} extended the approach of \citet{Rosenbaum1983} to a normal outcome and allowed for a continuous stratification variable by specifying the treatment assignment model. \citet{carnegie2016assessing} extended \citet{Rosenbaum1983}'s model to a continuous treatment and a normally distributed hypothesized unmeasured confounder $U$. Approaches taken in \citet{Rosenbaum1983}, \citet{Imbens2003}, and \citet{carnegie2016assessing} all specify parametric models that relate the outcome and the treatment assignment to observed covariates and a scalar hypothesized unmeasured confounder. \citet{Dorie2016} proposed to more flexibly model the response surface using Bayesian Additive Regression Trees, while still keeping the parametric specification of the treatment assignment model.

\section*{A.2: Difference between our approach and \citet{Ichino2008}}
\begin{figure}[h]
    \centering
    \includegraphics[scale = 0.6]{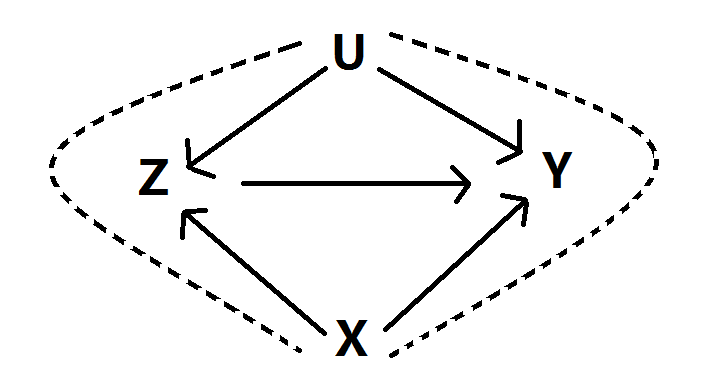}
    \caption{Illustration of the collider bias as in \citet{Ichino2008}}
    \label{fig: illustrate collide bias ichino}
\end{figure}

\citet{Ichino2008} considered a sensitivity analysis framework for propensity-score-matching-based estimators. Their sensitivity analysis framework is based on the following assumption:
\begin{equation*}
    \label{assumption: ichino}
    P(U = 1 \mid Y = i, Z = j, \mathbf{X}) = P(U = 1 \mid Y = i, Z = j) = p_{ij}, ~i,j \in \{0, 1\},
\end{equation*}
where $U$ is a hypothesized binary unmeasured confounder, $Z$ a binary treatmetn, and $Y$ a binary outcome. Their assumption says that once conditional on the treatment $Z$ and outcome $Y$, the unmeasured confounder $U$ is independent of observed covariates $\mathbf{X}$. However, as pointed out by \citet{Hernan1999} and \citet{VanderWeele2008}, this assumption does not hold in general: when $(Z, Y)$ are causally descendant of $U$ and $X$, conditioning on $(Z, Y)$ would introduce association between $U$ and $\mathbf{X}$ (even if $U$ and $\mathbf{X}$ are marginally independent), a phenomenon commonly known as ``collider bias'' in the literature. Figure \ref{fig: illustrate collide bias ichino} illustrates this point. On the other hand, our factorization of the joint likelihood $(\mathbf{X}, U, Z, Y^{(0)}, Y^{(1)})$ into $f(\mathbf{X})\cdot f(U \mid \mathbf{X}) \cdot f(A \mid \mathbf{X}, U) \cdot f(Y^{(0)}, Y^{(1)} \mid Z, \mathbf{X}, U)$ does not introduce such a problem. 

\begin{center}
{\large\bf Supplementary Material B: Details on the EM Algorithm}
\end{center}

\section*{B.1: The EM Algorithm}
The EM algorithm (\citealp{Dempster1977}) has been a popular technique for handling missing data problems. The general framework consists of two steps. In the E step, we form the conditional expectation of the complete-data log-likelihood, where the conditional expectation is taken with respect to the current model parameters: 
\begin{gather}
\label{eqn: E step}
Q(\boldsymbol\theta'\mid\boldsymbol\theta) = \mathbb{E}_{\boldsymbol\theta}[\log \{g(\boldsymbol x\mid\boldsymbol \theta')\}\mid\boldsymbol x_{obs}, \boldsymbol \theta].
\end{gather}
In this expression, $g(\boldsymbol x) = g(\boldsymbol x_{obs}, \boldsymbol x_{mis})$ is the complete-data likelihood, $\boldsymbol \theta$ is the estimated parameters at the current iteration and our objective is to find parameters $\boldsymbol \theta'$ that maximize Equation \ref{eqn: E step}. This maximization step is known as the M step. 

In particular, in the presence of missing covariates, \citet{Ibrahim1990} used EM by the method of weights to compute MLE. His method applies to any parametric regression models, including GLMs, random-effects models, parametric and semiparametric survival models. For our purpose, we use the EM by the method of weights for the missing binary covariate $U$; however, the method of weights may be used for any categorical or continuous covariates in a similar fashion.

\section*{B.2: Details on the E-step and M-step}
We describe in details the E-step and M-step in our current problem. Recall the EM algorithm involves calculating the expected complete-data log-likelihood. Let $\boldsymbol x_\ell = (\boldsymbol x_{\text{obs},\ell}, \boldsymbol x_{\text{mis},\ell})$ denote the complete covariate data for each subject $\ell$, $\ell = 1,2, ..., N$. In the current case, $\boldsymbol x_{\text{mis},\ell} = u_{\ell}$ and $\boldsymbol x_\ell = (\boldsymbol x_{\text{obs},\ell}, u_\ell)$, i.e., the missing covariate in our model is the hypothesized unmeasured confounder $U$. We further denote the set of parameters of interest as $\boldsymbol\gamma = (\boldsymbol\kappa, \boldsymbol\psi, \sigma, \beta)$. Let $l(\boldsymbol\gamma\mid\boldsymbol x_\ell, z_\ell, y_\ell)$ denote the log-likelihood for the $\ell^{th}$ observation. Note according to our model specifications, the complete-data log-likelihood for subject $\ell$ factors nicely and is given by
\begin{equation}
l(\boldsymbol\kappa, \boldsymbol\psi, \sigma, \beta\mid\boldsymbol x_\ell, z_\ell, y_\ell) = l_{y_\ell|\boldsymbol x_\ell, z_\ell}(\boldsymbol\psi, \sigma, \beta) + l_{z_\ell|\boldsymbol x_\ell}(\boldsymbol \kappa),
\label{eqn: complete-data log-lik}
\end{equation} where \[
l_{y_\ell|\boldsymbol x_\ell, z_\ell}(\boldsymbol\psi, \sigma, \beta) = \log\frac{1}{\sqrt{2\pi}\sigma} -\frac{(y_\ell - a_\ell - \boldsymbol\psi^T\boldsymbol x_{\text{obs},\ell} - \delta u_\ell - \beta z_\ell)^2}{2\sigma^2}
\] and \[
l_{z_\ell|\boldsymbol x_\ell}(\boldsymbol \kappa) = z_\ell\log \pi_\ell + 
(1 - z_\ell)\log(1 - \pi_\ell)
\]
with \[
\pi_\ell = \frac{\exp (\boldsymbol\kappa^T\boldsymbol x_{\text{obs},\ell} + \lambda u_\ell) }{1 + \exp(\boldsymbol\kappa^T\boldsymbol x_{\text{obs},\ell} + \lambda u_\ell)}.
\]
We can write the conditional expectation of the complete-data log-likelihood for subject $\ell$ as follows 
\begin{gather}
\label{eqn: Q_i}
Q_\ell = \sum_{u_\ell} l(\boldsymbol\gamma \mid \boldsymbol x_{\text{obs},\ell}, u_\ell, z_\ell, y_\ell)~p(u_\ell\mid\boldsymbol x_{obs, \ell}, y_\ell, z_\ell, \boldsymbol\gamma^{(s)}),
\end{gather} 
where $\boldsymbol\gamma^{(s)}$ denotes the current model parameters $(\boldsymbol\kappa^{(s)}, \boldsymbol\psi^{(s)},\sigma^{(s)}, \beta^{(s)})$ and the summation is over all possible realizations of the missing covariate $U$. In the binary case, the summation is simply over $u_\ell = 0$ and $1$. 

To maximize Equation \ref{eqn: Q_i}, we need to compute $p(u_\ell\mid\boldsymbol x_{obs, \ell}, y_\ell, z_\ell, \boldsymbol\gamma^{(s)})$, the posterior distribution of $U$ for subject $\ell$, given the treatment $z_\ell$, the observed covariates $\boldsymbol x_{\text{obs},\ell}$, the response $y_\ell$, and the current model parameters $\boldsymbol\gamma^{(s)}$. In the treated group, i.e., $Z_\ell = 1$, we have 

\begin{equation}
\begin{split}
\label{prob: posterior_full}
&P(U_\ell = 1 \mid \boldsymbol x_{\text{obs},\ell}, y_\ell, z_\ell = 1; ~\boldsymbol\gamma^{(s)}, ~p, \lambda,\delta) \\ 
=& \frac{p \cdot \text{expit}\{ {\boldsymbol\kappa^{(s)}}^{T}\boldsymbol x_{\text{obs},\ell} + \lambda\} \cdot \phi\big( \frac{r_\ell}{\sigma^{(s)}}\big)}{p \cdot \text{expit}\{ {\boldsymbol\kappa^{(s)}}^T\boldsymbol x_{\text{obs},\ell} + \lambda\} \cdot \phi\big( \frac{r_\ell}{\sigma^{(s)}}\big) + (1-p) \cdot \text{expit}\{ {\boldsymbol\kappa^{(s)}}^T\boldsymbol x_{\text{obs},\ell}\} \cdot \phi\big(\frac{r_\ell + \delta}{\sigma^{(s)}}\big)},
\end{split}
\end{equation}
where \[
r_\ell = y_\ell - a_\ell - {\boldsymbol\psi^{(s)}}^T \boldsymbol x_{\text{obs},\ell} - \delta - \beta^{(s)}.
\]
We have a similar expression for $P(U_\ell = 1)$ in the control group. 

For a fixed set of sensitivity parameters $(p, \lambda, \delta)$, the M step now reduces to finding, 
$\boldsymbol\gamma = (\boldsymbol\kappa, \boldsymbol\psi,\sigma, \beta)$ that maximizes:
\begin{equation}
\begin{split}
Q = \sum_{\ell = 1}^N Q_\ell &= \sum_{\ell = 1}^N \sum_{j = 0}^1 p(U_\ell = j\mid \boldsymbol x_{\text{obs},\ell}, y_\ell, z_\ell; \boldsymbol\gamma^{(s)})~l(\boldsymbol\gamma \mid \boldsymbol x_{\text{obs},\ell}, U_\ell = j, y_\ell, z_\ell) \\
&= \sum_{\ell = 1}^N \sum_{j = 0}^1 w_{\ell j} \big(l_{y_\ell | \boldsymbol x_\ell, z_\ell}(\boldsymbol\psi, \sigma, \beta) + l_{z_\ell|\boldsymbol x_\ell}(\boldsymbol\kappa)\big)\\
&= \sum_{\ell = 1}^N \sum_{j = 0}^1 w_{\ell j} ~l_{y_\ell | \boldsymbol x_\ell, z_\ell}(\boldsymbol\psi, \sigma, \beta) + \sum_\ell \sum_{j = 0}^1 w_{\ell j} ~l_{z_\ell | \boldsymbol x_\ell}(\boldsymbol\kappa)\\
&= Q^{(1)} + Q^{(2)},
\end{split}
\end{equation}
where \[
w_{\ell j} = P(U_\ell = j\mid \boldsymbol x_{\text{obs},\ell}, y_\ell, z_\ell; \boldsymbol\gamma^{(s)})
\] and is given by Equation \ref{prob: posterior_full}. Note to maximize $Q$, it suffices to maximize $Q^{(1)}$ and $Q^{(2)}$ separately. This reduces to finding the MLE for a weighted regression and a weighted logistic regression, which can be easily implemented in commonly used statistical softwares.

\begin{center}
{\large\bf Supplementary Material C: Additional Details on Reporting Sensitivity Analysis Results, Application, and Software}
\end{center}

\section*{C.1: Treatment effect heterogeneity}
Model (2) assumes a constant additive treatment effect $\beta$. Below, we describe a possible way to extend the method to heterogeneous treatment effect.

Consider a model that allows treatment heterogeneity:
\[
Y_{ij} \mid \boldsymbol x_{ij}, u_{ij}, z_{ij} = a_i + \boldsymbol\psi^T \boldsymbol x_{ij} + \delta u_{ij} + \beta_{ij} z_{ij} + \epsilon,
\]
where $\beta_{ij}$ is the treatment effect for subject $ij$, and $\epsilon \sim N(0, \sigma^2)$. Let $\beta = \mathbb{E}[\beta_{ij}]$, and we can rewrite the above model as the following:
\begin{equation}
    \begin{split}
Y_{ij} \mid \boldsymbol x_{ij}, u_{ij}, z_{ij} &= a_i + \boldsymbol\psi^T \boldsymbol x_{ij} + \delta u_{ij} + \beta z_{ij} + \underbrace{(\beta_{ij} - \beta) z_{ij} + \epsilon}_{\epsilon'}\\ 
&= a_i + \boldsymbol\psi^T \boldsymbol x_{ij} + \delta u_{ij} + \beta z_{ij} + \epsilon',
    \end{split}
\end{equation}
where $\epsilon' \sim N(0, \sigma^2_{ij})$ is now a possibly heteroskedastic error, and $\beta$ can be thought of as representing the average treatment effect. One can then model $\epsilon'$ to account for heterogeneous treatment effect. For instance, one may believe that the treatment effect is homogeneous within each matched set, but heterogeneous across different matched set. In this case, one may model $\epsilon' \sim N(0, \sigma^2_{i})$, and estimate $\sigma_i$ in addition to matched-set-fixed-effects $a_i$. A very similar EM algorithm would apply.

\section*{C.2: More results on treatment effect versus different ($p$, $\lambda$, $\delta$) pairs}
Table \ref{tbl: our_result addition in app} summarizes the estimated treatment effect versus additional (p, $\lambda$, $\delta$) pair. 

\begin{table}
\caption{\label{tbl: our_result addition in app} Treatment effect versus different (p, $\lambda$, $\delta$) pairs}
\centering
\fbox{%
\begin{tabular}{ccccc}
\hline
                & p = 0.5          & p = 0.3    & p = 0.1  & \multicolumn{1}{c}{\text{Simultaneous}}\\
\hline
($\lambda$, $\delta$) & 95\% CI    & 95\% CI    & 95\% CI   & \multicolumn{1}{c}{p-value}\\ \hline
(0, $\infty$)      & (0.402, 0.812)   & (0.401, 0.806) &(0.408, 0.819) & 0\\
($\infty$, 0)      & (0.399, 0.805)   & (0.398, 0.811) &(0.406, 0.810) & 0\\
(0.5, 5.0)      & (-0.548, -0.067) & (-0.095, 0.281)& (0.342, 0.724) &6.16e-6\\
(1.0, 5.0)      & (-1.365, -0.869) & (-0.756, -0.004) & (0.186, 0.580)&1\\
(5.0, 0.5)     &  (0.080, 0.502)& (0.172, 0.515) & (0.375, 0.719) &1\\
(5.0, 1.0)     & (-0.289, 0.086)   & (-0.234, 0.159) &(0.207, 0.585) & 1 \\
\hline
\end{tabular}}
\end{table}

\section*{C.3: Four snapshots of the interactive calibration plot}
\label{app: subsec: 4 snapshots}
\begin{figure}[H]
\centering
\caption{\label{fig: calibrate plot label}
Calibration plots with different $(\lambda, \delta)$ pairs for a fixed $p = 0.5$}
\includegraphics[width = \textwidth]{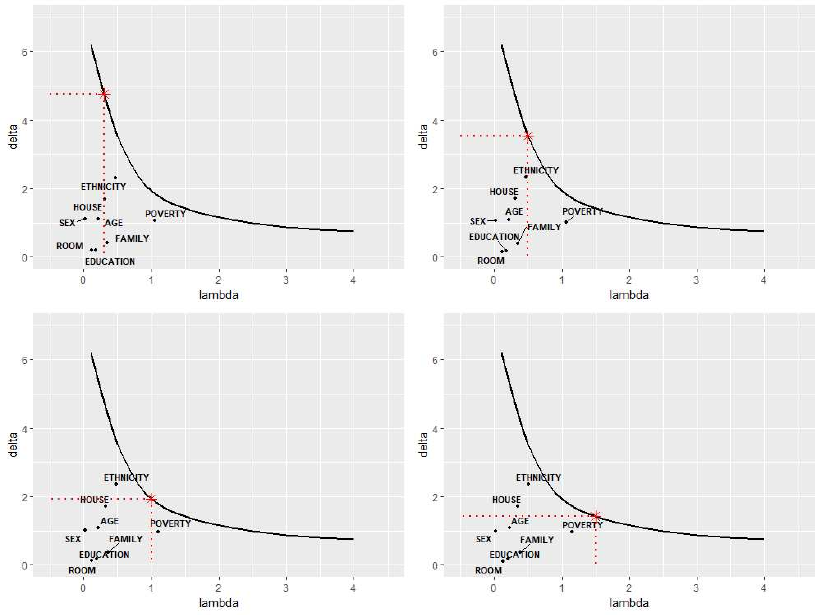}
\end{figure}

\end{document}